# Rapid feasibility assessment of components formed through hot stamping: A deep learning approach

Hamid Reza Attar[a], Haosu Zhou[a], Alistair Foster[b], Nan Li[a],*

*a Dyson School of Design Engineering, Imperial College London, London SW7 2DB, UK*
*b Impression Technologies Ltd, Coventry CV5 9PF, UK*

*Corresponding author
E-mail address: n.li09@imperial.ac.uk (N.Li)

## Abstract

The novel non-isothermal Hot Forming and cold die Quenching (HFQ®) process can enable the cost-effective production of complex shaped, high strength aluminium alloy panel components. However, the unfamiliarity of designing for the new process prevents its widescale adoption in industrial settings. Recent research efforts focus on the development of advanced material models for finite element simulations, used to assess the feasibility of new component designs for the HFQ® process. However, FE simulations take place late in design processes, require forming process expertise and are unsuitable for early-stage design explorations. To address these limitations, this study presents a novel application of a Convolutional Neural Network (CNN) based surrogate as a means of rapid manufacturing feasibility assessment for components to be formed using the HFQ® process. A diverse dataset containing variations in component geometry, blank shapes, and processing parameters, together with corresponding physical fields is generated and used to train the model. The results show that near indistinguishable full field predictions are obtained in real time from the model when compared with HFQ® simulations. This technique provides an invaluable tool to aid component design and decision making at the onset of a design process for complex-shaped components formed under HFQ® conditions.

## 1    Introduction

The growth in demand for lightweight materials in the automotive and other transport industries has accelerated in recent years, in response to the ever-growing concerns surrounding global environmental impact [1]. Aluminium alloys are a family of lightweight alloys with exceptional strength to weight ratios, beginning to find widespread applications in the automotive industry. Despite many advantages compared to conventional heavier steels, aluminium alloys exhibit poor formability under cold working conditions and display a significant degree of springback once removed from the forming tools [2].

To overcome these limitations, the novel Hot Forming and cold die Quenching (HFQ®) process has been developed by Lin *et al.* [3]. A summary of the HFQ® process is shown in Figure 1, and described further by Mohamed *et al.* [4]. Briefly, aluminium blanks are heated to their solution heat treatment (SHT) temperature, before being formed and quenched simultaniously in cold dies, followed by artificial aging. Since the material formability is improved under the designed thermal conditions [4], the HFQ® process enables enhanced design freedom from high strength aluminium alloys, at the production efficiency of a hot stamping process.

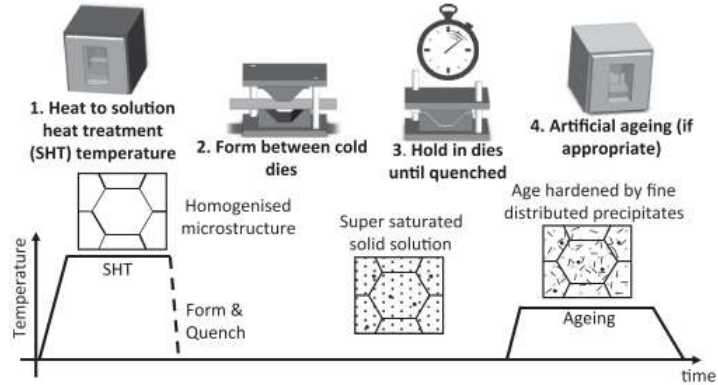

Figure 1 A schematic summary of the HFQ® process, adapted from Raugei *et al.* [1].

The HFQ® forming process is non-isothermal, due to the temperature difference between the heated blank and cold dies [5]. The material flow stress response of aluminium alloys at elevated temperatures is temperature and strain rate dependant [4,6]. In hot stamping, the temperature and strain rate vary dynamically with both time and location in the sheet metal, which is unlike traditional and more familiar cold stamping processes [7]. Due to these inherent manufacturing complexities, there is a lack of familiarity amongst industrial designers. Consequently, it has been observed that the uptake of HFQ® technology is limited in modern industrial settings, meaning the potential weight saving and design capabilities of the new process are currently not realised to the full potential.

Published literature on HFQ® focuses largely on the development of advanced material models that predict the material constitutive behaviour under HFQ® conditions [4,6,8–10]. These models are then used to accurately simulate the forming response of HFQ® formed components using Finite Element (FE) simulations [5,9,11] in order to assess their manufacturing feasibility. Though useful, forming simulations usually take place late in design processes, when the component design is near completion. Further, the numerical cost and forming process expertise deem FE simulations unsuitable for early stage decision making and design exploration, particularly for the new, less familiar HFQ® process.

The advent of machine learning (ML) has introduced a new way to establish time efficient process models. ML, and in particular supervised learning, approaches fundamentally comprise of learning system dynamics from representative input-target data samples in a process known as training. Once trained, the ML models can be employed to infer system behaviours when presented with previously unseen inputs [12]. Such models are commonly known in the literature as surrogate models.

Surrogate models based on FE simulation data have been extensively used to support the design of sheet forming processes. Classic approaches involve scalar based models to regress the performance landscape, given a low dimensional design space parameterisation. As typical examples, Harsch *et al.* [13] used surrogate models to construct process forming window maps for a cold metal stamping process. Ambrogio *et al.* [14] employed a Kringing surrogate model technique to establish a relationship between incremental sheet forming processing parameters and final sheet thickness. However, those models are only effective for scenarios that can be parameterised using relatively few parameters and are highly dependent on a well-designed parameterisation schema. Examples of such scenarios are process parameter optimisation assuming a fixed geometry [15] or relatively simple shapes [16]. Engineering companies may conduct hundreds of forming simulations each day, resulting in an accumulation of large datasets that may not share common CAD parameterisations, especially for complex geometries that are often described by several dimensions. Data that falls outside of the pre-determined parameterisation therefore cannot be leveraged by conventional approaches.

In addition to the above limitations, the data extracted from simulations is commonly simplified into a manufacturing feasibility scalar evaluator, such as maximum strain. As mentioned by Zhou, Li & Xu [17], scalar based models lack informativeness due to this data consolidation. This is particularly true in scenarios where the manufacturing feasibility is determined by distribution based indicators, such as post form thickness gradients [18], surface slip lines [19] and wrinkle distributions [20]. To provide richer data representations, other references predicted full field data on a FE mesh using deep neural networks (DNNs). For example, Sauer, Schleich & Wartzack [21] trained a DNN model to predict the 10,000 node-based equivalent plastic strain for a locking tooth component and Pfrommer *et al.* [22] used DNNs to predict shear angles from over 24,000 elements in textile forming simulations. However, these models predict the mesh dependent simulation results; the element type, numbering strategy and mesh connectivity must be common between the training data and the model predictions. A mesh independence is desired for generalisation in forming applications, since components are of different shapes and sizes.

To overcome the aforementioned limitations, researchers have recently borrowed modelling techniques from the field of deep learning, a subset of machine learning. In particular, Convolutional Neural Networks (CNNs), in conjunction with image representations of design inputs and computer simulation results outputs have been employed to preserve prediction informativeness without a mesh based dependency [17,23–29]. CNNs are a particular class of neural networks that have gained popularity when working with spatially structured data such as grids or images. Nie, Jiang & Kara [30] used CNNs to predict stress fields in 2D cantilever structures deforming elastically. Thuerey *et al.* [31] predicted full flow fields around varying aerofoil shapes at several flow conditions. Lino *et al.* [32] and Fotiadis *et al.* [33] simulated the dynamics of wave propagation within varying curved and multi-faceted open and closed geometries.

The above studies have proven that CNNs provide strong predictive capabilities when modelling highly non-linear physical systems. However, the applications of CNNs to predict the manufacturing feasibility of components formed through hot stamping has not yet been explored. The complex non-isothermal nature of the HFQ® process together with the strain-rate dependency of aluminium alloys at elevated temperatures [4,6] make the process a challenging physical system to accurately model using conventional surrogate techniques.

The purpose of this study is to present a novel application of a CNN based surrogate model as a means of real time manufacturing feasibility assessment of components to be formed through the HFQ® process. To the best of the authors' knowledge, this is the first time CNN based deep learning techniques are applied to a non-isothermal forming process. The main contributions of this work are listed below:

- A novel method for processing CAD geometries and FE simulation results from stamping simulations into image based representations for CNN training and evaluation.

- A new dataset for corner shaped components formed under non-isothermal HFQ® conditions.

- The implementation of the CNN-based approach for rapid manufacturing feasibility assessment, given a component geometry, blank shape, and processing parameters.

- The demonstration of the approach for the non-isothermal HFQ® process, where complex forming responses such as deformation, thinning localisation and wrinkling are predicted.

- A comparison between the predicted and real FE simulated forming responses, establishing the effectiveness of the approach.

## 2 Overview of the proposed component feasibility assessment approach

An overview of the proposed approach is presented in Figure 2. Typically, a conventional feasibility assessment for a production process is carried out following the simplified process shown in the green boxes. The feasibility of candidates design is assessed by forming simulations based on finite element analysis (FEA) [5]. As mentioned in the introduction, forming simulations are unsuitable for design explorations and early stage decision making. Since the simulator would have to be run by a process engineer with knowledge of the HFQ® process each time a designer wishes to make a geometry change, this makes the design process slow and costly. A typical engineering approach would therefore be to only simulate a small set of potential design candidates without an extensive search of potential design solutions. This process can be prolonged by improper initial designs for HFQ® by designers unfamiliar with its process limits. Consequently, the HFQ® process may be overlooked by designers and more conventional less efficient, heavier, or higher embedded $CO_2$ solutions may be adopted instead.

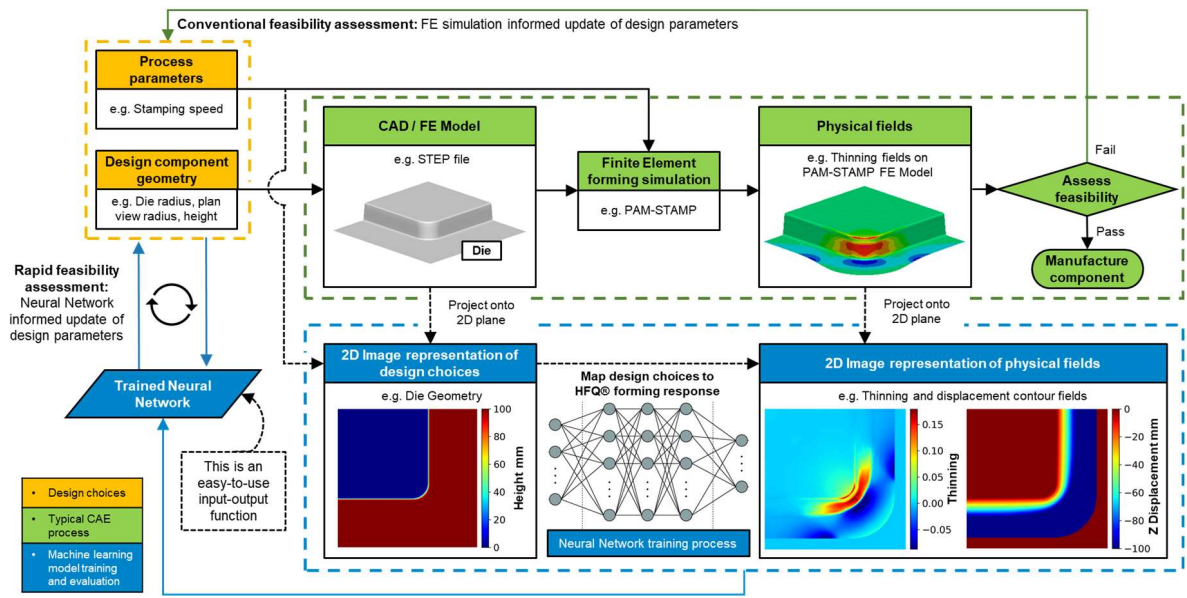

Figure 2 An overview of the proposed design feasibility assessment approach (blue boxes) for HFQ®, compared with the conventional approach (green boxes).

This paper presents a data-driven alternative approach to rapidly predict component manufacturing feasibility of geometries to be formed using HFQ® technology without extensive knowledge of the process, for use at the onset of a design process, seen by the blue boxes in Figure 2. Data from design choices (e.g., component geometry, etc) together with forming simulation results are first mapped into a suitable form before being used to train a CNN based surrogate model. Once trained, the surrogate will be used to assess the feasibility of new component geometries to be HFQ® formed in real time. Once suitable solutions are found from the model, detailed component design can continue, with only manufacture-intent geometries reviewed by the FE simulator, thus removing the need for extensive FEA iterations. Therefore, this paper provides an invaluable design support tool, enabling designers to become familiar with the design capability of the HFQ® process and design without requiring the involvement of an HFQ® expert.

As well as variations in geometry, the proposed CNN based surrogate model takes as input HFQ® forming process parameters such as blank shape, which a designer can approximate using common CAD unfolding tools, and manufacturing parameters. The outputs are taken to be the displacement and thinning fields, which are used to evaluate the manufacturing feasibility of the designed component.

# 3 Problem definition and design of experiments

Prior to model training, a diverse dataset must be established which captures the key forming characteristics of the considered system. As mentioned in the overview, data used to train the model is sourced from FE forming simulations. The problem definition, details on the design of experiments are discussed in this section.

## 3.1 Geometry and blank definition

### 3.1.1 Geometry

Being common limiting design features for rectangular or square components, deep drawn corners are considered in this study as an example use case. Specifically, flanged shrink corners are considered, which are formed when blank material is drawn into a plan view radius that is smaller than the blank corner radius, as described by Horton *et al.* [34]. Due to their symmetry, quarter boxes are modelled, with a half side length of 500 mm. Three corner fillet radii and the deep draw height are considered as geometric parameters, as seen in Figure 3 together with some example die geometries. The tool geometries are determined by surface offsets from designed components.

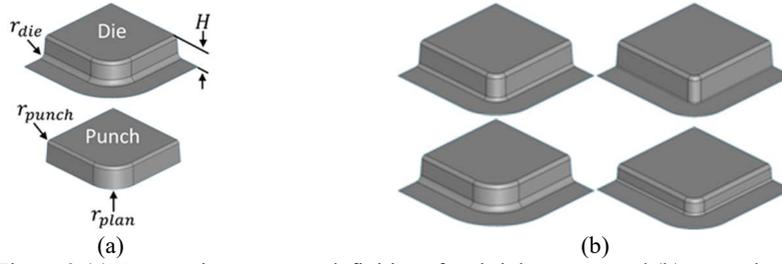

Figure 3 (a) Geometric parameter definitions for shrink corners and (b) example die geometries.

### 3.1.2 Blank shape

The blank shape has been reported in the literature to be a critical factor in ensuring a successfully formed component under HFQ® conditions. Politis *et al.* [11] explained that blank shape optimisation was essential in forming trials of a complex shaped deep drawn door inner geometry using HFQ® conditions. Zhu *et al.* [35] demonstrated how local necking was eliminated when removing excessive blank material from a L-shaped deep drawn component. Since part designers are often not familiar with creating optimal blank shapes, it is useful to include the effect of blank shape, while understanding that the final blank design will be optimised at a later stage of development.

The blank shapes in this study are schematically in Figure 4. Figure 4(a) shows a section through a generic as-formed corner and Figure 4(b) gives the undeformed blank dimensions. The blank half-length and blank radius are denoted by $L_{blank}$ and $r_{blank}$ respectively in the figure and are functions of the component dimensions to ensure sufficient blank material. The flange width and punch half width are denoted by $F$ and $W$ and are kept constant at 50 and 500 mm, respectively. To allow further variation in the blank shapes, the blank dimensions are scaled by scaling factors $A$ and $B$, as seen in Figure 4(b).

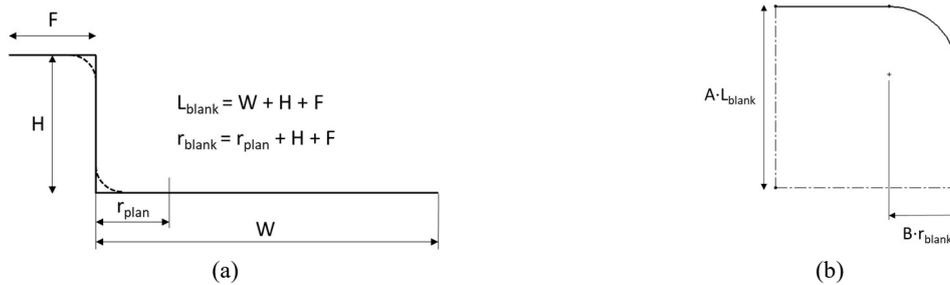

Figure 4 Determination of undeformed blank shape geometries with (a) section through the as-formed corner with calculation of blank half length $L_{blank}$ and blank radius $r_{blank}$, and (b) square quarter blank with scaling factors $A$ and $B$.

## 3.2 HFQ® Simulation setup

To build the simulation models, the tool and blank CAD variant generation was automated using the VBA programming language in SolidWorks. Filtering rules as well as upper and lower bounds were applied to preserve geometric integrity during the automation process, as described by Ramnath *et al.* [36,37]. To mesh the CAD geometries for the FE simulations, the commercial FE pre-processing software HyperMesh was used. Within HyperMesh, the TCL programming language was utilised to automate the FE mesh generation.

Due to its specialisation for sheet metal forming, the FE software PAM-STAMP was utilised for the HFQ® forming simulations. Due to the vast number of simulations required to populate the dataset, the Python programming language was used in conjunction with PAM-STAMP's scripting interface to load in the prepared mesh files for each geometry and automate the computations. Each simulation model represents one unique sample in the dataset and consists of five components, as shown by the example in Figure 5. Due to the symmetry of the problem, only one quarter of the geometry was modelled, and symmetry boundary conditions were imposed.

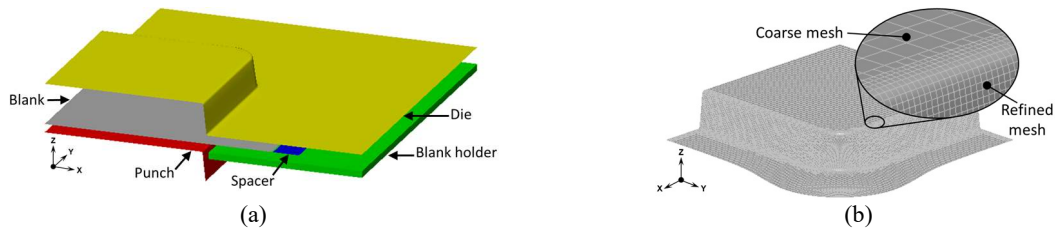

Figure 5 Finite element model example: (a) cross sectional view and (b) HFQ® formed component.

To simulate non-isothermal HFQ® conditions, coupled thermo-mechanical simulations were conducted. During each simulation, the die and blank holder were kept at a fixed distance apart, determined by the spacer thickness. The punch was fixed in all degrees of freedom, while the die, spacer and blank holder could move in the vertical direction under imposed displacement boundary conditions to deform the blank over the punch. The tools were modelled as mechanically rigid bodies but with active temperature degrees of freedom to allow for heat transfer. The blank was modelled using deformable reduced integration Belytschko-Tsay shell elements with a 2 mm blank thickness for all samples, and five through thickness integration points. The main process and simulation parameters are given in Table 1 and the physical properties of the blank and tools are given in Table 2, based on typical values found in HFQ® literature [4,5,38,39].

Table 1 Main process and simulation parameters.

| Parameter | Value |
| --- | --- |
| Initial workpiece temperature °C | Variable – See Table 3 |
| Initial tooling temperature °C | 25 |
| Stamping speed mm/s | Variable – See Table 3 |
| Spacer thickness mm | Variable – See Table 3 |
| Friction coefficient (all interfaces) | 0.1 [38] |
| Velocity scaling factor | 10× |

Since a large number of simulations were needed, efforts were made to reduce simulation time where possible, while maintaining the integrity of the simulation results. Adaptive mesh refinement was used on an initial course blank mesh to refine mesh iteratively during the computation only where it is needed, for example, to capture local bending characteristics, as highlighted in Figure 5(b). A velocity scaling factor was used to further reduce the simulation times, and the strain rate correction feature in PAM-STAMP was enabled to maintain the correct strain rates at scaled stamping speeds. Simulation times were on average 5 minutes for each sample on an Intel Core i7-9700 CPU using 8 processors.

Table 2 Physical properties of the AA6082 blank and tool steel.

| Property | AA6082 | Tool steel |
|---|---|---|
| Thermal conductivity (tonne·mm)/(s$^3$·K) | 170 | 20 |
| Specific heat capacity (mm$^2$)/(s$^2$·K) | 8.9E8 | 5E8 |
| Density tonne/mm$^3$ | 2.7E-9 | 7.8E-9 |
| Poisson's ratio | 0.33 | 0.3 |
| Young's modulus MPa | 7E4 | 2.1E5 |
| Heat transfer coefficient | Function of gap and pressure [40] | |

## 3.3 Material model

To capture the material constitutive behaviour under HFQ® conditions, the FE simulations used a temperature and strain rate dependant viscoplastic material model for AA6082, adapted from El Fakir [40]. This material model enables the viscoplastic effects on the flow stress of AA6082 after SHT at elevated temperatures to be considered and covers typical temperature and strain rate ranges seen during the HFQ® forming process. Examples of flow stress-strain curves plotted from the material model are shown in Figure 6. Further details of the material model can be found in the original paper [40].

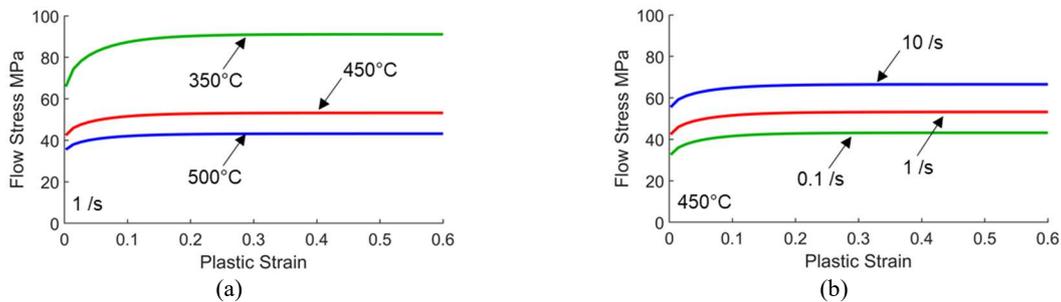

Figure 6 Temperature and rate dependant flow stress-strain model [40] for AA6082 used in HFQ® simulations.

## 3.4 Preliminary HFQ® processing simulations

Recall that the purpose of the numerical experiments was to establish a diverse dataset that captures all the key forming characteristics of the considered system. As such, the most important factors which influence the material forming response were first identified by conducting preliminary simulations. The influence of geometric parameters when forming corners was previously characterised by the authors of the present paper [41]. The effect of processing parameters is now considered.

With the AA6082 material model introduced above, a representative corner geometry was formed under a combination of initial blank temperatures and stamping speeds under HFQ® conditions. Figure 7 highlights variation in post-form thinning as a consequence of varying these processing parameters. Under these conditions, it was found that a lower temperature and higher forming speed contributed to improved thickness uniformity. This trend is in good agreement with the conclusions drawn by Zheng *et al.* [42] when experimentally investigating the hot drawability of AA6082 under HFQ® conditions.

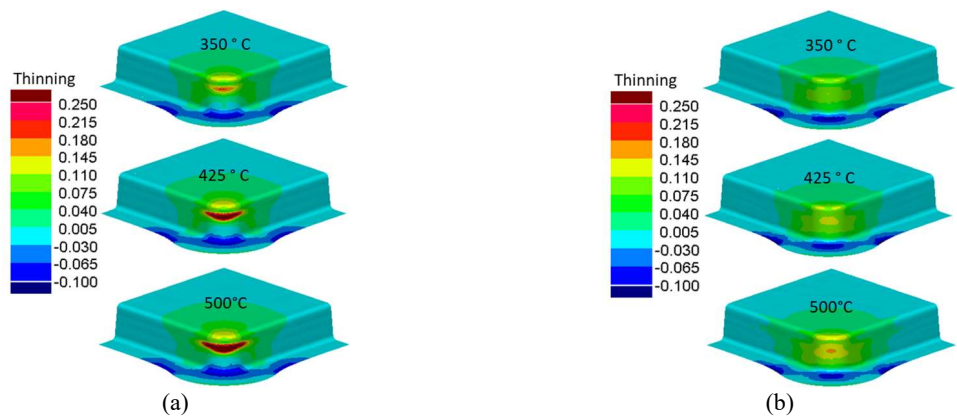

Figure 7 Thinning response of a typical AA6082 HFQ® formed shrink corner at initial blank temperatures of 350, 425 and 500 °C at speeds of (a) 50 mm/s and (b) 500 mm/s.

Forming with an initial blank temperature of 350 °C at 500 mm/s produced an optimum forming response, however these processing conditions might not be easily achievable. Since in a HFQ® process, is a solution heat treatment step (at 525 °C for AA6082 [43]) is required prior to forming [3,4], pre-cooling to 350 °C might not always be possible and may require specialist equipment. Further, existing presses used for older and more conventional forming technologies, such as cold forming or hot stamping steels, may operate at lower maximum press speeds than the speeds typically required for HFQ® (250 mm/s and above [44]). For example, Ganapathy *et al.* [45] use speeds as low as 60 mm/s when hot stamping a boron steel B-pilar component. Therefore, there is a need to evaluate a candidate component design under a range of processing parameters, to determine whether or not a feasibility window exists, using available resources, for that geometry when formed using the HFQ® process.

To further exploit the predictive capability of the model, samples with wrinkling were deliberately included in the dataset. In this study, wrinkling was provoked by varying the spacer thicknesses (see Figure 5(a)). The effect of varying the spacer thickness can be seen in Figure 8 where large ripples can be seen in the flange area at greater spacer thicknesses. It was also found that wrinkles induce regions of localised thinning on the lower sidewall of the corner component, as seen in the figure, due to the increased difficulty to draw in material.

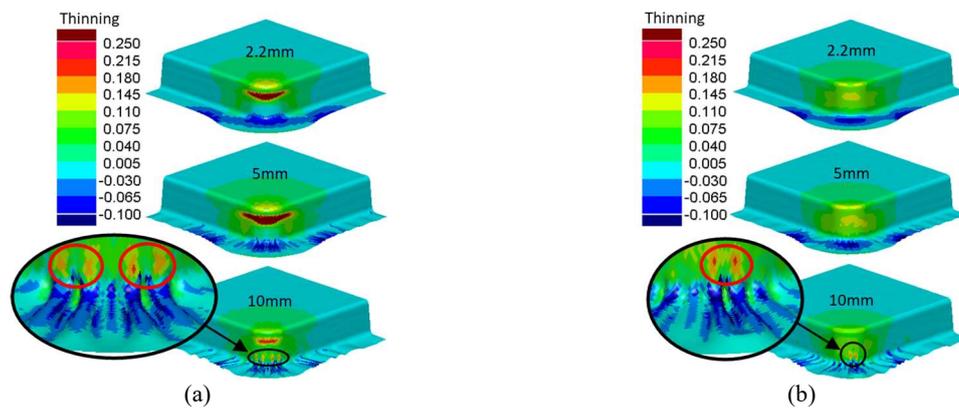

(a)                                                                 (b)

Figure 8 Thinning and wrinkling response of a typical AA6082 HFQ® formed shrink corner with spacer thicknesses of 2.2, 5 and 10 mm at speeds of (a) 50 mm/s and (b) 500 mm/s.

Following the preliminary forming simulations, the novel HFQ® dataset developed in this study included the effects of initial blank temperature, stamping speed and spacer thickness, to capture their associated effects. Another merit of including such parameters into the dataset is that the developed model is compatible with engineering data collected over time by engineering companies, which may be generated under different process settings.

## 3.5    Design of experiments

The parameters considered in this study are summarised in Table 3. Each parameter was assigned a defined range to generate a diverse distribution of forming characteristics within the dataset. Examples of forming characteristics to be included in the dataset include excessive localised thinning due to non-uniform temperatures distributions and sharp tool geometric features, thickening, and excessive wrinkling.

It is to be noted that extreme forming responses such as excessive thinning, and wrinkling are indeed undesirable when forming a real component, and are considered as forming induced defects [4,15]. However, only by including samples (i.e., combinations of input variables) which are responsible for producing such unwanted forming induced defects can the model learn to understand when these defects occur. Therefore, the parameter ranges were deliberately selected to induce such defects onto some samples. As an example, the spacer thickness upper bound was set to 10 mm, meaning wrinkles with a 10 mm ripple height would be induced for those samples as seen in Figure 8.

Table 3 Considered parameters and ranges.

| Parameter | Symbol | Group | Range |
| --- | --- | --- | --- |
| Die radius | $r_{die}$ | Geometry | $5 - 25$ mm |
| Punch radius | $r_{punch}$ | Geometry | $5 - 25$ mm |
| Plan view radius | $r_{plan}$ | Geometry | $60 - 120$ mm |
| Design height | $H$ | Geometry | $60 - 120$ mm |
| Blank width scaling factor | $A$ | Blank shape | $1.1 - 0.9$ |
| Blank radius scaling factor | $B$ | Blank shape | $1.1 - 0.1$ |
| Spacer thickness | $t_s$ | Processing | $2 - 10$ mm |
| Initial blank temperature | $T_{init}$ | Processing | $350 - 500$ °C |
| Stamping speed | $S$ | Processing | $50 - 500$ mm/s |

In total, 1800 unique samples, consisting of CAD geometries, undeformed blank shapes, and processing parameters, were built. The Latin hypercube [46] design of experiments (DoE) technique was employed, together with the variable ranges given in Table 3. It is a popular sampling strategy for deterministic computer simulations, enabling a uniform distribution of samples within the design space while avoiding recurrences.

# 4    Data preparation

The component geometries, blank shapes, processing parameters and corresponding FE simulation results are used for training and evaluating the CNN based surrogate model. Before training, the data must be mapped into a unified image representation. Details on the novel data preparation strategy used in this study are given in this section.

## 4.1    Input images

For any stamping process, an undercut-free geometry is a necessary requirement in avoiding collision with forming tools. As such, the 3D die geometries were projected onto a 2D plane (i.e., an image) with a height contour without information loss. All images in this study used uniform Cartesian grids with a fine resolution of $256 \times 256$ pixels to capture small geometric features in the die, such as sharp radii. To perform the projection, the CAD models were first converted into meshes, which enabled them to be represented by a discrete set of nodes. The mesh nodes were treated as point clouds with Cartesian coordinates in 3D space and the out of plane height values were then interpolated onto a uniform 2D Cartesian grid.

Similarly, the blank position was also interpolated onto an image to form a binary map, with a pixel value of '1' where there is material and '0' for void regions. It is conjectured in this paper that representing scalar inputs as images improves model prediction performance, since it allows the CNN kernels to convolve on those images and learn the corresponding kernel weights during training. To generate an image representation of the scalar process parameters, the spacer thickness $t_{s_i}$, initial blank temperature $T_{init_i}$ and stamping speed $s_i$ for sample $i$ were all multiplied to '1' and '0' blank images to form image representations. Consequently, the pixel location on these images remained a representation of the blank geometry, while the pixel values represented the scalar parameters. An example sample from the dataset is shown in Figure 9.

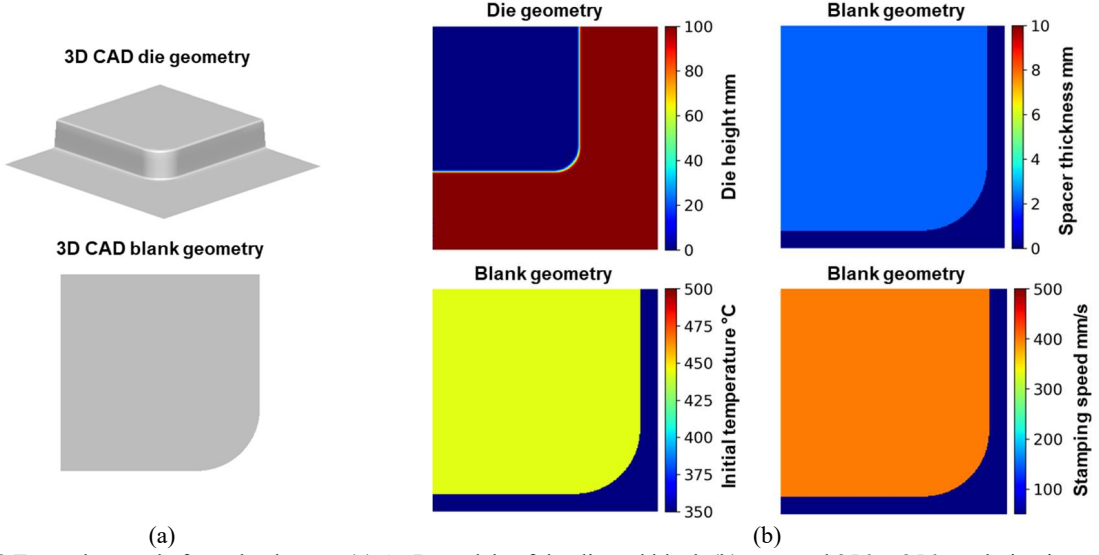

(a)                                                                              (b)

Figure 9 Example sample from the dataset: (a) CAD models of the die and blank (b) prepared $256 \times 256$ resolution input images for network training and evaluation.

The procedure for obtaining input images from CAD files is summarised in Algorithm 1. Once collated, one sample from the dataset had an input size of $4 \times 256 \times 256$. It is to be noted that Algorithm 1 can be implemented in any programming language and the meshing steps performed using open source meshing libraries, such as PyMesh [47], for a fully automated procedure.

---

**Algorithm 1** Procedure for obtaining input images for network training and evaluation

---

**Require**: $n_s$ = total number of samples, $H \times W$ = image resolution, $\boldsymbol{G}$ = Die CAD geometries, $\boldsymbol{B}$ = Blank CAD geometries, $\boldsymbol{t_s}$ = Spacer thicknesses, $\boldsymbol{T_{init}}$ = Initial temperatures, $\boldsymbol{S}$ = Stamping speeds

**Arrays**:

- $\boldsymbol{G} = [g_1, g_2, \ldots, g_{n_s}]$ where $g_i$ is the die CAD geometry of sample $i$

- $\boldsymbol{B} = [b_1, b_2, \ldots, b_{n_s}]$ where $b_i$ is the blank CAD geometry of sample $i$

- $\boldsymbol{t_s} = [t_{s_1}, t_{s_2}, \ldots, t_{s_{n_s}}]$ and $t_{s_i}$ is the spacer thickness of sample $i$

- $\boldsymbol{T_{init}} = [T_{init_1}, T_{init_2}, \ldots, T_{init_{n_s}}]$ and $T_{init_i}$ is the initial blank temperature of sample $i$

- $\boldsymbol{S} = [s_1, s_2, \ldots, s_n]$ and $s_i$ is the stamping speed of sample $i$

1: **for** $i = 1, \ldots, n_s$ **do**
2:     Sample $g_i$, $b_i$, $t_{s_i}$, $T_{init_i}$, and $s_i$ from prepared arrays
3:     Die mesh ← die CAD geometry $g_i$
4:     Grid heights ← interpolate(die nodal heights, $H \times W$ grid) // 1 die image
5:     Blank mesh ← blank CAD geometry $b_i$
6:     Grid blank positions ← interpolate(blank nodal positions, $H \times W$ grid)  // 1 blank binary map
7:     Multiply scalars $t_{s_i}$, $T_{init_i}$, and $s_i$ individually to blank binary map // 3 scalar parameter blank images
8:     Channel wise concatenation of images from 4. and 7. // obtain $4 \times H \times W$ array for sample $i$
9:     Store sample $i$ from 8. in $n_s \times 4 \times N \times N$ array
10: **end for**
11: Export pre-processed input $n_s \times 4 \times N \times N$ array

---

## 4.2   Simulation quality assessment

Before processing the simulation results into target images, the quality of the FE simulation results was assessed to ensure a high quality dataset. Extreme variable sets (e.g., excessively tight radii and large spacer thicknesses) were included in the dataset to provoke forming induced defects as mentioned previously. Because of this, it was found that some samples in the dataset contained physically implausible numerical outliers on the FE simulation results. Figure 10 shows two examples of physically implausible outliers on simulation results caused by material splitting due to excessive thinning, and material folding due to the drawing of wrinkles. In both cases, element distortions were found, with approximately 52% thinning and 75% thickening respectively, which clearly indicate material failures, but those values are not be physically plausible. These samples had to be corrected since they risked deteriorating the performance of the model.

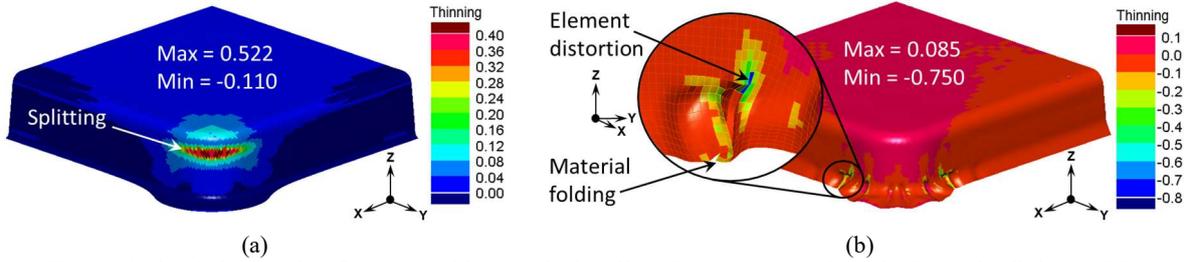

Figure 10 Simulation results of samples with numerical outliers due to (a) excessive thinning and splitting and (b) excessive wrinkling.

The distributions of samples with such outliers were clipped at their 99.5th percentiles to eliminate the risk of dataset corruption. A sample was flagged during image pre-processing as having implausible outliers if the maximum thinning or thickening values across its entire distribution violated predefined thresholds. Thresholds $c_1$ and $c_2$ were defined as maximum allowable thinning and thickening (i.e., min allowable thinning) respectively. The distributions of the flagged samples were then clipped and an example of thinning spectrums across all elements before and after clipping a sample with excessive wrinkling are shown in Figure 11. This approach was found to preserve the quality of the dataset.

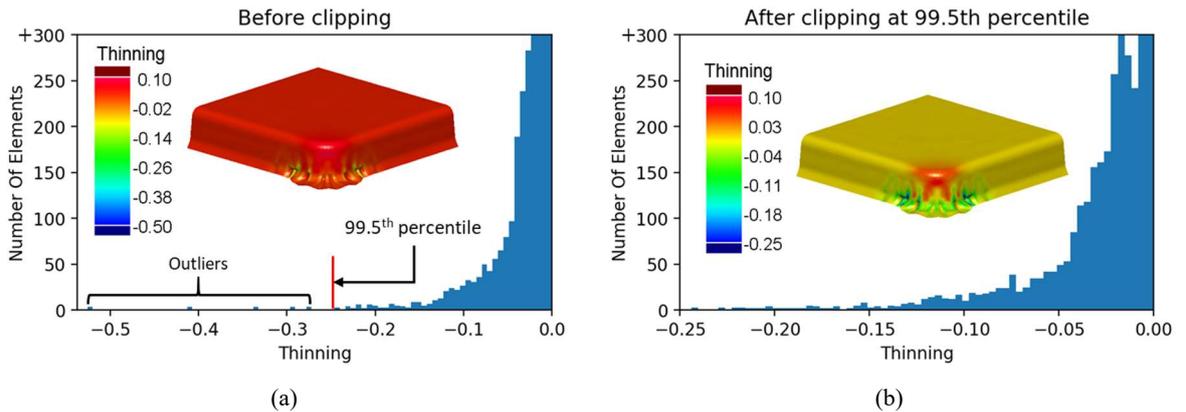

Figure 11 Example of distributions on a sample with wrinkling: (a) before and (b) after clipping at 99.5th percentile.

## 4.3   Target images

The target images are defined as the ground truths (i.e., actual FEA observations) for the network predictions to target during training. In this paper, these are images processed from the HFQ® simulation results. The simulation results considered were the post form thinning fields as well as X, Y, and Z components of displacement fields.

Similar to the input images, FE nodes were taken as point clouds and interpolation operations are used to convert nodal data to images. However, the thinning fields are computed at the element integration points, rather than at nodes, by the FE solver. To avoid troublesome calculation of the locations of integration points, the elemental field data was converted to nodal data, by averaging the elemental values to the connected nodes, as graphically illustrated in Figure 12. In the figure, $t_i^{(A)}$ is the thinning value from sample $i$ at element $A$, and $t_{i,k}$ is the thinning value at node $k$ averaged over its connected elements. The required connectivity information was extracted from the FE mesh files. Reduced integration elements were used for the simulations in this study, and therefore each element had only one integration point at the mid thickness.

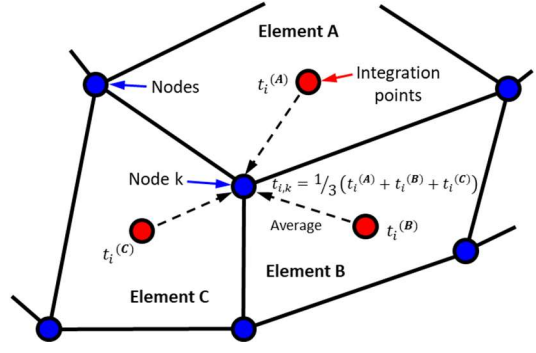

Figure 12 Illustration of averaging elemental field values to connected nodes.

Once the elemental field values were converted to nodal field values, the nodal point cloud of the as-formed component was converted back into the undeformed blank shape to enable a convenient 2D configuration for the images. Since adaptive mesh refinement was used during the FE computation (see Figure 5(b)), the nodes at the final stamping stage (which contain the field values) were different to the initial set of nodes belonging to the undeformed blank. Equation (1) was used to obtain the undeformed positions from the nodes at the final stamping stage,

$$\boldsymbol{d}_{0,i} = \boldsymbol{d}_i - \boldsymbol{\delta}_i \tag{1}$$

where $\boldsymbol{d}_{0,i}$ are the 3D nodal coordinates of the undeformed flat blank, $\boldsymbol{d}_i$ are the 3D nodal coordinates of the as-formed component and $\boldsymbol{\delta}_i$ is the 3D nodal displacement field calculated by the FE simulation, all for sample $i$. The 2D point cloud data from the undeformed flat blank was then interpolated onto a uniform Cartesian grid, and field values were masked to '0' where there is no blank material using the blank shape binary maps (see Section 4.1). This process is illustrated in Figure 13, and the overall proposed procedure for obtaining target images is summarised in Algorithm 2, which again is automatable.

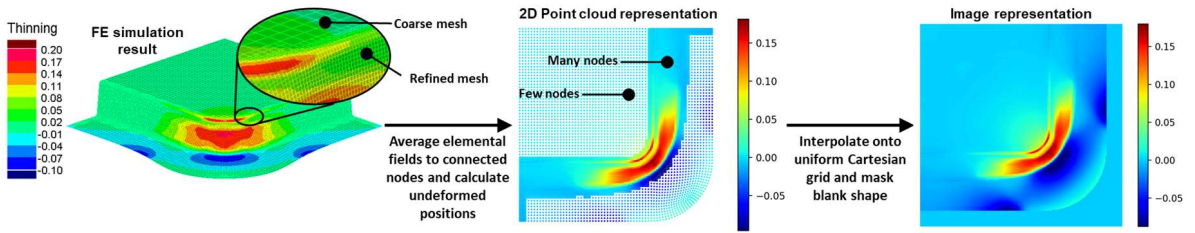

Figure 13 Illustration of converting an FE simulation result into an image representation with thinning contour displayed.

---

**Algorithm 2** Procedure for obtaining target images for network training and evaluation

---

**Require**: $n_s$ = total number of samples, $n_f$ = total number of fields, $H \times W$ = image resolution, $\boldsymbol{c}$ = allowable extreme elemental values, $\boldsymbol{t}$ = thinning fields, $\boldsymbol{\delta}$ = displacement fields, $\boldsymbol{d}$ = deformed nodal coordinates, $\boldsymbol{C}$ = element connectivity

**Arrays**:

- $\boldsymbol{c} = [c_1, c_2]$, where $c_1$ and $c_2$ are maximum and minimum allowable elemental values respectively

- $\boldsymbol{t} = [\boldsymbol{t_1}, \boldsymbol{t_2}, \; ..., \; \boldsymbol{t_{n_s}}]$ where $\boldsymbol{t_i}$ is a vector containing the elemental thinning field of sample $i$

- $\boldsymbol{\delta} = [\boldsymbol{\delta_1}, \boldsymbol{\delta_2}, \; ..., \; \boldsymbol{\delta_{n_s}}]$ where $\boldsymbol{\delta_i}$ is a vector of the 3D nodal displacement field of sample $i$

- $\boldsymbol{d} = [\boldsymbol{d_1}, \boldsymbol{d_2}, \; ..., \; \boldsymbol{d_{n_s}}]$ where $\boldsymbol{d_i}$ is a vector of the 3D deformed nodal coordinates of sample $i$

- $\boldsymbol{C_e} = \left[ \boldsymbol{C_{e_1}}, \boldsymbol{C_{e_2}}, \; ..., \; \boldsymbol{C_{e_{n_s}}} \right]$ where $\boldsymbol{C_{e_i}}$ is the element connectivity of sample $i$, commonly found in any standard FE mesh file

1:   **for** $i = 1, ..., n_s$ **do**
2:       Sample fields $\boldsymbol{t_i}$, $\boldsymbol{\delta_i}$, $\boldsymbol{d_i}$ and $\boldsymbol{C_{e_i}}$ from prepared arrays
3:       $\boldsymbol{d_{0,i}} \leftarrow \boldsymbol{d_i} - \boldsymbol{\delta_i}$ // undeform final stage nodes
4:       Determine nodal connectivity $\boldsymbol{C_{n_i}}$ from $\boldsymbol{C_{e_i}}$
5:       **for** $j = 1, ..., n_f$ **do**
6:          **if** field $j$ is element based **do** // i.e., if field was calculated at integration points by FE solver
7:             field $j$ ← average(field $j$, $\boldsymbol{C_{n_i}}$) // average field to connected nodes using nodal connectivity $\boldsymbol{C_{n_i}}$
8:             **if** max(field $j$) > $c_1$ **or** min(field $j$) < $c_2$ **do**
9:                field $j$ ← clip(field $j$, 99.5$^{th}$ percentile) // clip extremes to eliminate outliers
10:             **end if**
11:         **end if**
12:          field $j$ ← interpolate(field $j$, $\boldsymbol{d_{0,i}}$, $H \times W$ grid) // interpolate onto a uniform Cartesian grid
13:          Store field $j$ from 12. in $n_f \times H \times W$ array
14:       **end for**
12:       Store sample $i$ from 13. in $n_s \times n_f \times H \times W$ array
15:   **end for**
16:   Export pre-processed target $n_s \times n_f \times H \times W$ array

---

# 5   Network architecture and training

A neural network architecture was developed to be tailored for extracting characteristic features of the input images and mapping those features to corresponding HFQ® simulation results. Details of the network architecture and employed training scheme are given in this section.

## 5.1   Neural network architecture

The fundamental components of the neural network architecture utilised in this paper are its convolutional layers. By convolving learnable kernels across input images, convolutional neural networks (CNNs) hieratically detect and extract locational features. Important design features on a component geometry or blank shape were therefore automatically detected without human supervision, deeming CNNs suitable for the application of component feasibility analysis. The network architecture is presented in Figure 14.

A Res-SE-U-Net was selected in this paper to serve as the CNN architecture. Res-SE-U-Nets have recently been proven to be effective architectures for image-to-image translation tasks by a number of studies [17,25,30]. This novel architecture consists of two key components: a U-Net backbone with concatenative skip connections and integrated ResNet layers with Squeeze and Excitation modules (Res-SE) (detailed in Figure 15). U-Nets were originally proposed by Ronneberger, Fischer & Brox [48] for the application of biomedical image feature extraction and segmentation. Since then, U-Nets have been extensively applied to model physical systems, such as simulating wave dynamics by Lino *et al.* [32] and predicting airflows around aerofoils by Thuerey *et al.* [31] to name a few.

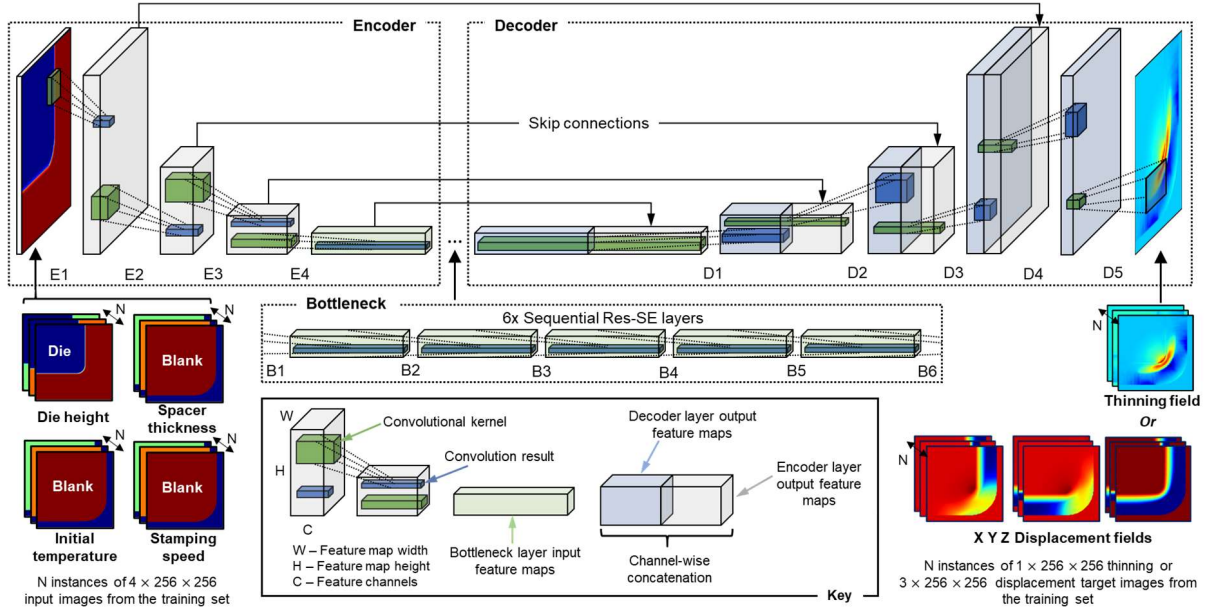

Figure 14 Res-SE-U-Net architecture of the neural network: a U-Net structure with Res-SE architectural blocks at its bottleneck, and illustration of images used for network training.

A U-Net consists of an encoder-decoder set up with concatenative skip connections established to bridge intermediate feature maps. The contracting path (encoder) and expansive path (decoder) can be seen in the left and right side of Figure 14 respectively. At each down-sampling operation in the encoder, the image spatial dimensions are reduced while the number of feature channels are increased. The encoder outputs a spatially condensed representation of the input images at the bottleneck of the architecture. Intuitively, this dense representation contains a "summary" of the most important features contained in the inputs. The dense representations are then passed through the bottleneck (detailed below) before being fed into the decoder. The decoder is a mirror of the encoder; at each up-sampling (i.e., up convolution) operation in the decoder, the image spatial dimensions are increased while the number of feature channels are reduced. The novel skip connections in the U-Net architecture perform channel-wise concatenations of feature maps from the encoder to the decoder. This ensures feature reusability and thereby helps recover spatial information lost during down-sampling [48]. Apart from the final layer, all convolutional layers in the encoder and decoder were followed by Batch Normalisation (BN) and ReLU layers. Table 4 summarises the configuration details of the network layers used in this study.

Table 4 Dimensionality and configuration of each layer in the designed Res-SE-U-Net.

| Layer | Function | Input channels | Output channel | Spatial kernel | Stride | Padding | BN? | Activation Function |
|---|---|---|---|---|---|---|---|---|
| Encoder | | | | | | | | |
| E1 | Conv2d | 4 | 16 | $9 \times 9$ | 1 | 4 | Yes | ReLU |
| E2 | Conv2d | 16 | 32 | $8 \times 8$ | 2 | 3 | Yes | ReLU |
| E3 | Conv2d | 32 | 64 | $6 \times 6$ | 2 | 2 | Yes | ReLU |
| E4 | Conv2d | 64 | 128 | $4 \times 4$ | 2 | 1 | Yes | ReLU |
| Bottleneck (6x identical Res-SE layers with SE block reduction ratio $r = 16$) | | | | | | | | |
| B1 to B6 | - | 128 | 128 | | See Figure 15 | | | |
| Decoder | | | | | | | | |
| D1 | ConvTransposed2d | 256 | 64 | $4 \times 4$ | 2 | 1 | Yes | ReLU |
| D2 | ConvTransposed2d | 128 | 32 | $6 \times 6$ | 2 | 2 | Yes | ReLU |
| D3 | ConvTransposed2d | 64 | 16 | $8 \times 8$ | 2 | 3 | Yes | ReLU |
| D4 | Conv2d | 32 | 8 | $5 \times 5$ | 1 | 2 | Yes | ReLU |
| D5 | Conv2d | 8 | 1 or 3 | $5 \times 5$ | 1 | 2 | No | None |

The bottleneck, defined as the zone between the encoder and decoder, is responsible for processing spatially dense representations (i.e., smallest H and W dimension) of the inputs, converting them into dense representations of the simulation results. Since dense representations are processed in this region of the architecture, Res-SE layers were chosen to enhance information flow. A Res-SE layer consists of a Residual Network (ResNet) module [49] with an integrated Squeeze and Excitation (SE) block [50]. Information is processed by the Res-SE layers as shown by the flow diagrams Figure 15.

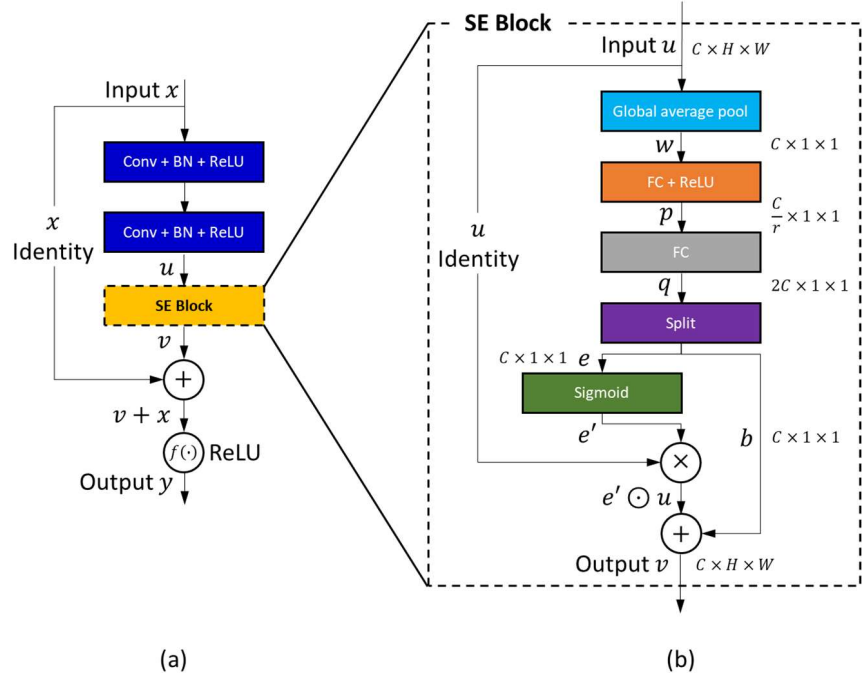

Figure 15 Composition of a Res-SE layer: (a) ResNet module with characteristic skip connection and integrated SE block shown in (b).

The utilisation of ResNet module introduces several benefits in terms of model performance. Unlike the concatenative skip connections used in the U-Net backbone, the additive skip connection shown in Figure 15(a) enables an alternative identity mapping between the input and output of a ResNet module to be established. The identity mapping helps preserve gradients to combat the vanishing gradient problem [49]. Further, the identity mapping also helps tackle the degradation problem, where network *training* accuracy has been reported to drop with increasing the number of layers [51,52]. This is because it is easier to optimise a residual mapping. As an extreme, when an identity is deemed optimal during training, it would be easier to push the weight layers (Conv + BN + ReLU and SE block layers in Figure 15) to zero rather than fitting an exact identity mapping, as first hypothesised by He *et al.* [49] and later confirmed by Li *et al.* [53]. Overall, the ResNet modules allow for deep neural architectures to be established, and the weight layer depths to be dynamically selected during training.

The SE blocks inside the ResNet modules are novel architectural units used as channel-wise attention mechanisms, first introduced by Hu, Shen & Sun [50]. To boost the representational capability of the network, the SE blocks adaptively recalibrate channel-wise feature maps within the ResNet module by explicitly establishing interdependencies between channels [50]. As shown in Figure 15(b), the inputs to the SE block $\boldsymbol{u} \in \mathbb{R}^{C \times H \times W}$ are first *squeezed* by a global average pool layer into a spatially averaged representation vector $\boldsymbol{w} \in \mathbb{R}^{C \times 1 \times 1}$, where $C$, $H$ and $W$ denote the channel, height and width dimensions. Then two fully connected (FC) layers are used to downscale $w$ with a reduction ratio $r$ (FC+ReLU layer) into $\boldsymbol{p} \in \mathbb{R}^{C/r \times 1 \times 1}$ and then upscale (second FC layer) into $\boldsymbol{q} \in \mathbb{R}^{2C \times 1 \times 1}$. To embody the channel-wise recalibration of the input $\boldsymbol{u}$, $\boldsymbol{q}$ is spit along the channel wise direction to form $\boldsymbol{e} \in \mathbb{R}^{C \times 1 \times 1}$ and $\boldsymbol{b} \in \mathbb{R}^{C \times 1 \times 1}$, where $\boldsymbol{e}$ is passed through a sigmoid activation function before scaling the input $\boldsymbol{u}$ while $\boldsymbol{b}$ provides an offset bias. The result of the SE block is the recalibrated or *excited* tensor $\boldsymbol{v} \in \mathbb{R}^{C \times H \times W}$.

## 5.2 Network training

A random 10% of the data was held back for testing while the remaining 90% was used for training. Each input instance was of size $4 \times 256 \times 256$, following Section 4.1. For the target images, the thinning images were of size $1 \times 256 \times 256$ and the displacement images were of size $3 \times 256 \times 256$, with the 3 representing the X, Y, and Z components of displacements. Separate Res-SE-U-Nets were trained on input-thinning and input-displacement data pairs to prevent the sharing of network weights, which would otherwise occur if both types of outputs were to be predicted simultaneously by a single network. These networks both followed an identical architecture to Figure 15 but varied only in the output channel size, as seen in the figure and in Table 4.

The Mean Square Error (MSE) loss function was selected to evaluate network performance during training. The network prediction $\tilde{y}$ and ground truth $y$ are both displayed as $256 \times 256$ images, and were reshaped into a 1D array of length $n = 256^2$ before being passed into Equation (2) to evaluate the MSE.

$$MSE = \frac{1}{n} \sum_{j=1}^{n} (y_j - \tilde{y}_j)^2 \qquad (2)$$

Through an iterative training process, the optimiser sought to find the combination of network parameters to minimise the MSE loss function. In this way, the network was able to learn a function that can predict the HFQ® simulation results, and therefore manufacturing feasibility, given a new component geometry and process set up.

The models were trained in the PyTorch framework using the commonly recommended Adam optimiser [54] with default beta parameters $\beta_1 = 0.9$ and $\beta_2 = 0.999$, learning rate of 0.0005 and batch size of 20. The training loss history is shown in Figure 16, where steady declines can be seen indicating training stability for the designed network architecture. Training was set to be terminated at 1500 epochs. The thinning network took 3.4 hours to achieve convergence on the test set (300 epochs) and displacement network took 11.3 hours (1000 epochs). Both networks were trained on a NVIDIA Quadro RTX 5000 GPU.

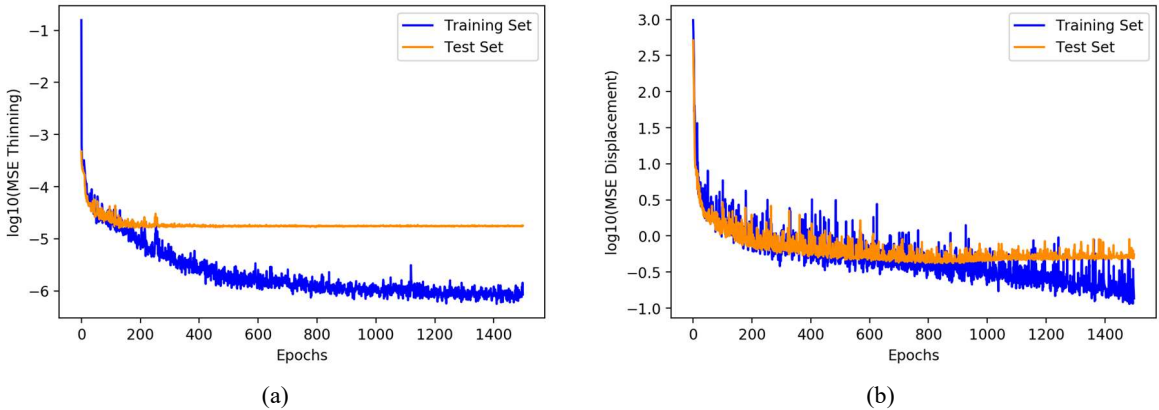

Figure 16 Loss histories of the proposed neural network models: (a) thinning network and (b) displacement network.

# 6 Results and discussion

The performance of the trained networks was evaluated to understand to what extent the models could be used for rapid HFQ® feasibility assessment. Details of the network performance, advantages and limitations are given in this section. For better reader visualisation, the scalar field distribution maps are presented as coloured images, note that the networks process monochrome images.

## 6.1 Thinning field prediction

The thinning field prediction performance was evaluated by comparing the predictions given by the network with ground truths (i.e., actual HFQ® FE simulation results). For brevity, network predictions are denoted as PD and ground truths as GT hereafter. Cases from the test set were used for evaluating the network predictions. These cases contained new component geometries formed under different processing conditions and different blank shapes to those seen by the network during training.

Figure 17 compares the thinning fields from five randomly sampled cases from the test set, without wrinkles, predicted from the network with equivalent ones from FE simulations. For all cases, indistinguishable distributions between network predictions and ground truths can be seen. Recall that the network does not solve any system dynamics iteratively like a FE solver does and has therefore learnt the physics of the HFQ® forming process from data supplied during training.

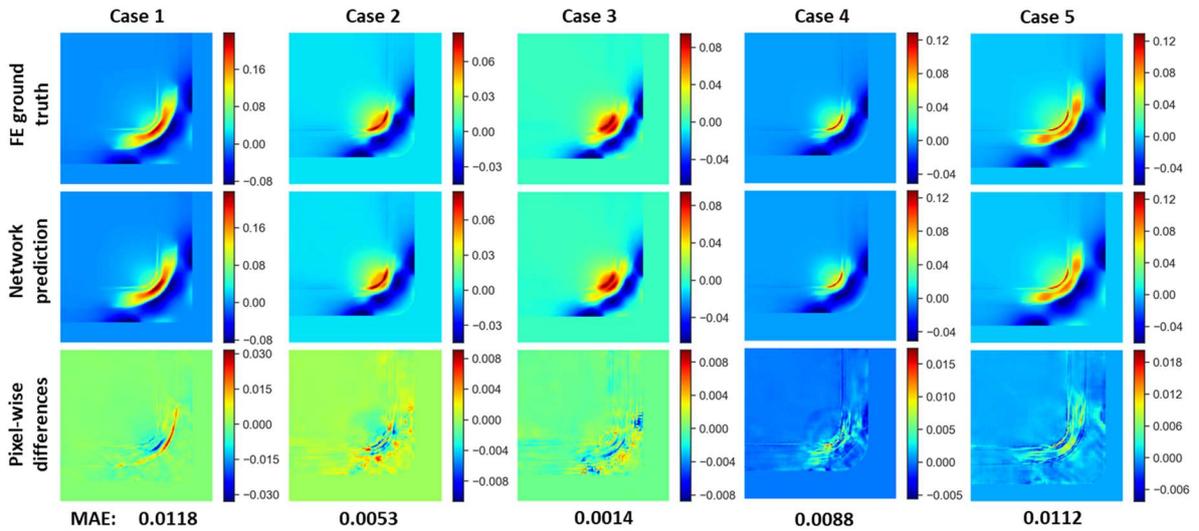

Figure 17 Comparison of ground truths with network predictions for thinning distributions and pixel wise differences for five random test set cases without wrinkles. Contours plotted on undeformed blank geometry.

A similar comparison can be seen in Figure 18 but this time for test set samples which showed the greatest maximum absolute error (MAE) between the network prediction PD and ground truth GT, defined as in Equation (3).

$$MAE = |max(PD) - max(GT)| \qquad (3)$$

It was found that the samples responsible for the greatest MAE for thinning prediction were samples which had experienced wrinkling when formed. In Figure 18, wrinkling is evidenced by the nebulous thinning distributions in *both* the PD and GT images, particularly around the blank borders, and were later confirmed by viewing the displacement distributions (discussed below). Such patterns were not present in the samples shown in Figure 17 which had no wrinkling and clear distributions.

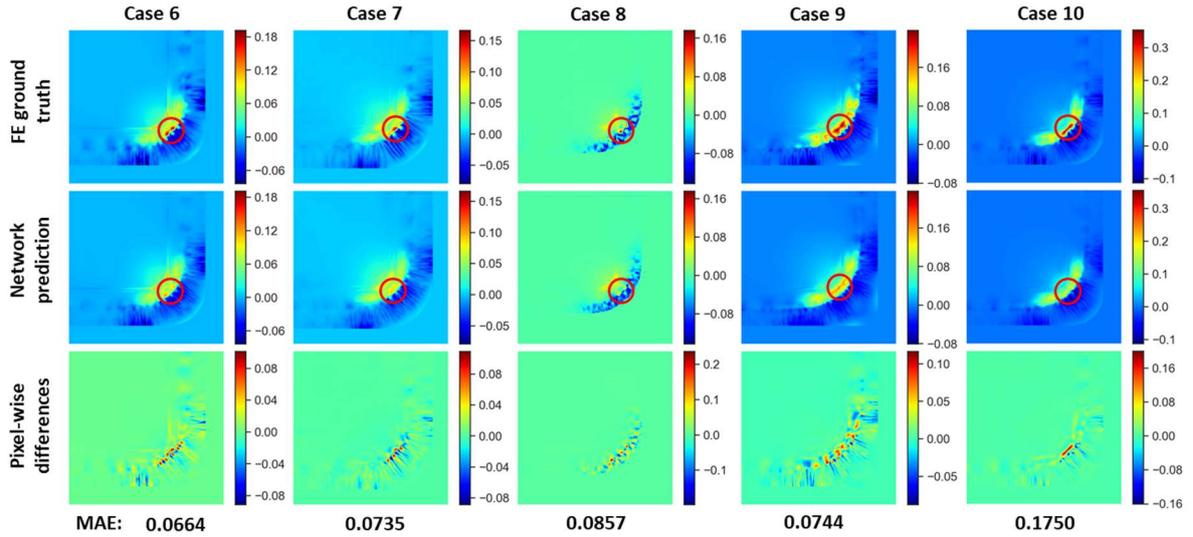

Figure 18 Comparison of ground truths with network predictions for thinning distributions and pixel wise differences for the top five largest MAE test set cases, all of which had wrinkles. Contours plotted on undeformed blank geometry. Red circles highlight small local differences in distributions.

To study the predicted thinning distributions in more depth, Case 1 from Figure 17 (without wrinkles) and Case 6 from Figure 18 (with wrinkles) are taken as representative samples. Figure 19 presents the thinning distributions cut along a diagonal line through these cases. For the case without wrinkles, an excellent agreement between PD and GT can be seen. The case with wrinkles still shows the characteristic peaks, although the peaks do not match exactly between the PD and GT.

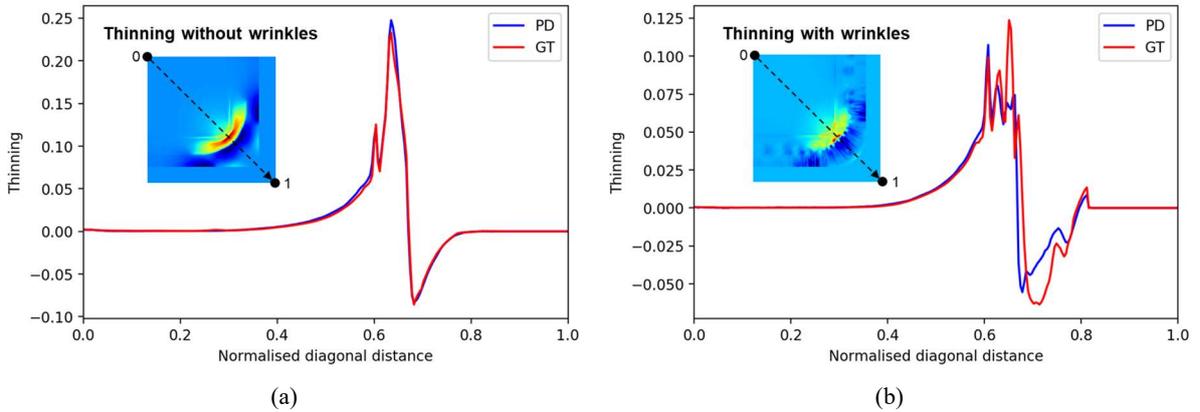

Figure 19 Thinning along a diagonal line for (a) Case 1 from Figure 17 (without wrinkles) and (b) Case 6 from Figure 18 (with wrinkles).

To further evaluate the overall thinning prediction performance, the data distributions are visualised by violin plots in Figure 20 and the Kullback-Leibler divergences (KLD) between the GT and PD distributions are calculated, shown in the figure headings. The KLD is selected, among others, as a metric to quantify the similarity between two distributions, where a value of 0 indicates identical distributions. To represent the distributions, the max and mean values of each image in the training and test sets were calculated for both GT and PD. As seen in Figure 20, all distributions show excellent agreement and are confirmed by the small values of KLD between GT and PD distributions.

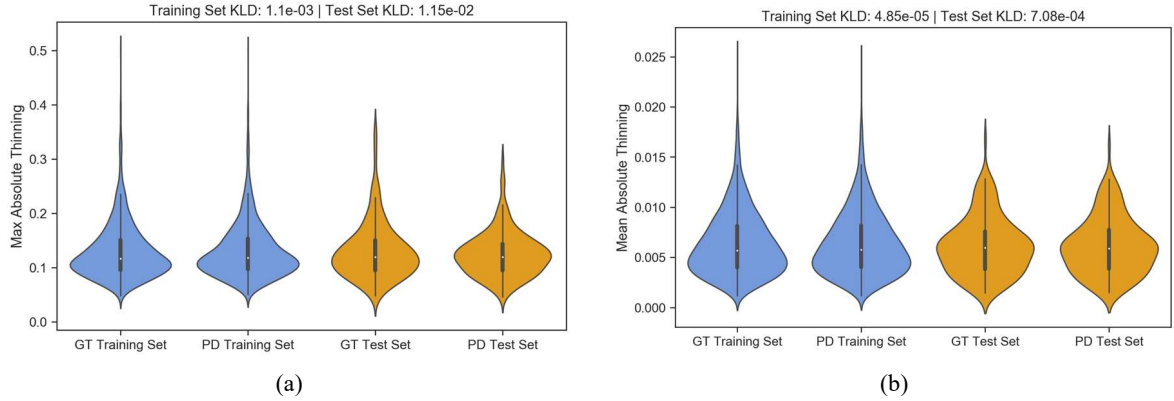

<div align="center">(a)            (b)</div>

Figure 20 Comparison of dataset distributions with KLD similarity metric between PD and GT, for training and test sets, for (a) max and (b) mean absolute thinning.

## 6.2 Displacement field and wrinkle prediction

A comparison between the X, Y, and Z displacement fields obtained from network predictions and ones from ground truths were made and presented in Figure A.1 in the Appendix. The cases shown are the five random test set cases from Figure 17 where displacement field network predictions are seen to be indistinguishable from their FE simulation counterparts.

The results presented thus far have been plotted on 2D undeformed blank shapes and were the raw outputs from the trained networks. The high quality displacement predictions exemplified in Figure A.1 enabled the displacement fields predicted by the network to be used to generate 3D as-formed predictions. By deforming a uniform cartesian grid using the 2D displacement vector fields (i.e., X, Y, and Z components of displacement for each pixel), 3D representations of the as-formed components were established. Thinning distributions were then superimposed as shown in Figure 21 for representative test case samples without and with wrinkles.

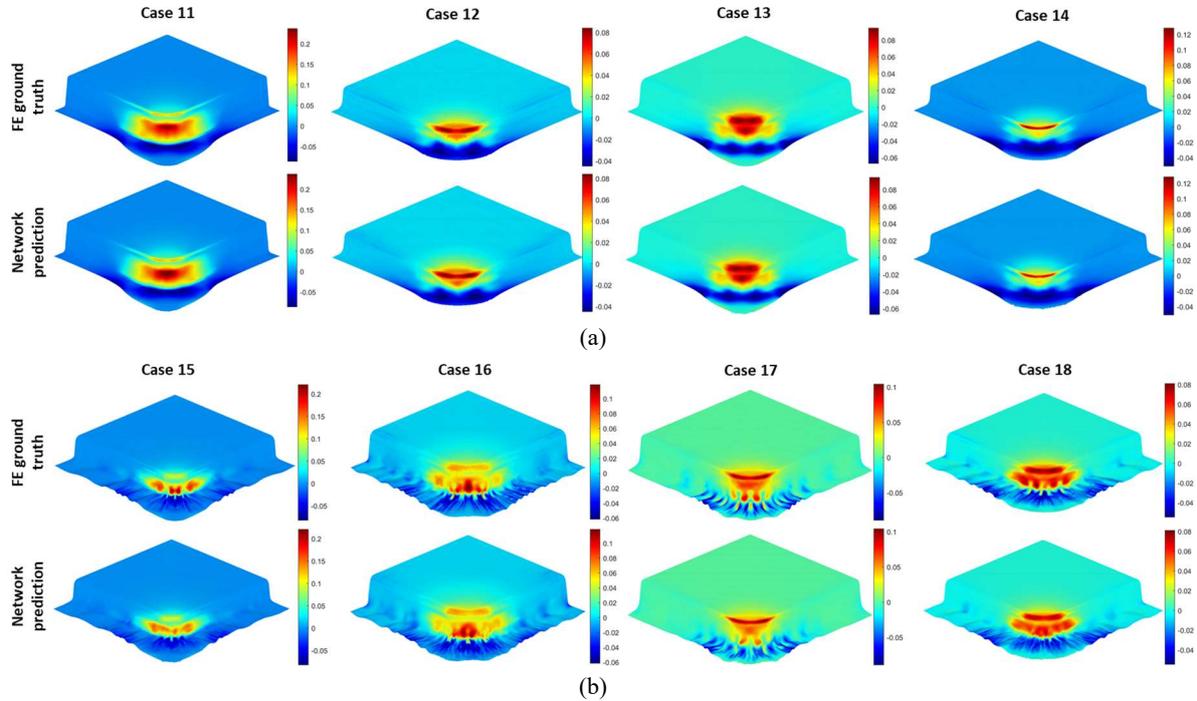

Figure 21 Comparison of ground truth and network prediction of 3D as-formed components with thinning distributions superimposed for samples (a) without and (b) with wrinkles. All cases from the test set, unseen during training.

To analyse the prediction of wrinkles further, Case 16 from Figure 21 is selected as a representative sample for a closer look, shown in Figure 22. Plotted are contours of wrinkle height, and a comparison between the ground truth and network prediction can be seen. It is seen that wrinkle heights are correctly predicted with remarkable accuracy. Specifically, ripple patterns and locations are also correctly predicted; there are zones of wider wrinkles near the straight edges of the as-formed component, while narrower wrinkles at the transition zone between the straight edges and the corner.

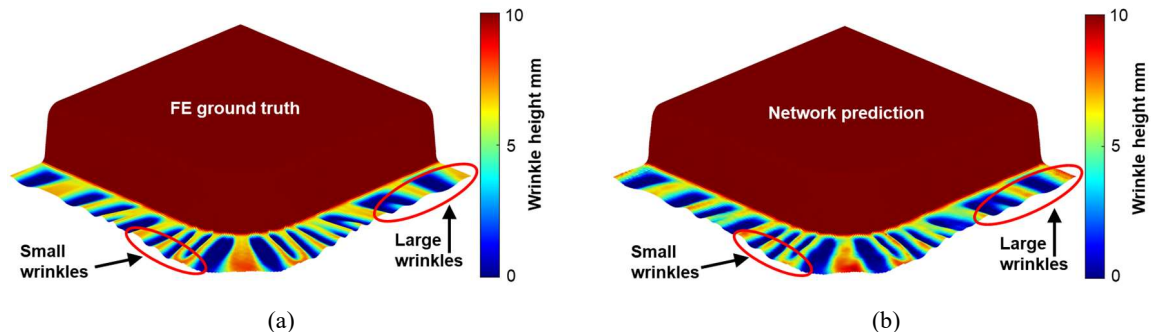

Figure 22 Comparison of wrinkle height prediction on the 3D as-formed component geometry for Case 16 from Figure 21: (a) ground truth and (b) network prediction.

## 6.3 Effect of training data size

The effect of training data size on the generalisation performance was investigated by evaluating the thinning field predictions by additional networks trained with smaller dataset sizes. Since unseen geometries are the most difficult factor to generalise to, the smaller datasets varied only in the number of die geometry images. To remove bias towards specific geometries, three random seeds were used in conjunction with the Latin Hypercube method to create three different datasets for each dataset size. For the performance evaluation, a test set containing 160 unique die geometries unseen to these networks during training was used, and MSE and mean relative error (MRE) metrics were calculated. Figure 23 shows the variations in MSE and MRE when varying dataset sizes.

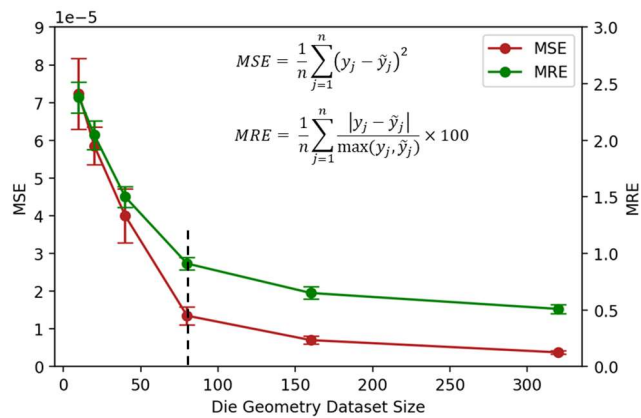

Figure 23 Effect of training data size on the thinning field network prediction performance on test sets.

Both the MSE and MRE declined when increasing dataset size. The standard deviation across the three datasets for each dataset size is plotted as the error bars, which also decrease with increasing dataset size, suggesting improved generalisation performance to unseen geometries. It is noteworthy that the trend is non-linear, and the decline is not as pronounced above a certain threshold; in this case, more than 80 unique die geometries in the dataset provided very little performance gain. Clearly the threshold will depend on the considered design space, but a general conclusion can be drawn that excessively large datasets might not be required to achieve acceptable generalisation performance for the proposed network, even under the complex non-isothermal HFQ® forming conditions.

## 6.4  Discussion

This work presents a technique for the rapid manufacturing feasibility evaluation of any geometry formed through the HFQ® process. Through the use of deep CNNs, thinning and wrinkling fields were predicted in real time, given a candidate geometry, blank shape and set of HFQ® processing parameters. By comparing the predicted fields with their FE simulation counterparts, near indistinguishable field outputs (i.e., forming responses) were obtained from the networks. Despite the accuracy of the predictions, a number of questions still need to be addressed.

The largest MAEs for the thinning distribution in the test set were found to be on wrinkled samples. Since wrinkle patterns are largely randomised, it was hypothesised that the reason the networks slightly underperformed in resolving local wrinkling induced thinning defects (see Figure 8) was because the networks were less exposed to such complex and disordered thinning patterns during training. However, since thinning defects are less important when wrinkles are present, this was not explored at length in this work, and wrinkle prediction took precedence. Nevertheless, excellent performance was seen overall.

The scope of the network was limited to components with standard shrink corner features in this study (Figure A.2(a)). The dataset developed can be extended to include more complicated geometric features. Some variants of complex shaped corners are shown in Figure A.2(b-c), commonly used on some battery tray designs [55]. A major strength of the image based technique used in this paper is that such complex geometries, typically represented by many CAD dimensions, can be uniformly represented as images.

To understand how the networks would process the more complex shapes in Figure A.2(b-c), intermediate feature maps from encoder layers E1 and E4 were visualised, shown in Figure A.3. Recall that the networks were trained only on images of standard corners (Figure A.2(a)). The feature maps reveal that the convolutional kernels learnt during training on standard corners are still able to detect the more complex stepped sidewall or chamfer corner design features. The encoder part of the network may interpret the shapes learnt from the standard corners as curves made up of lots of straight walls of size equal to the stride size. In this case, the higher the image resolution and smaller the stride size, the greater the encoder has tenancy to generalise to shapes outside the domain of shapes seen during training. This could mean that when new engineering data (i.e., from FE simulations) becomes available for new corner shapes, only the decoder part can be fine-tuned, i.e., by transfer learning, to capture the physical fields associated with the new geometries. Note that in an industrial setting, continual enrichment of training datasets with novel data samples would result in an increasingly proficient model which would require less and less retraining over time.

The methodology developed was able to predict the physical fields through being trained on examples seen in the training dataset, instead of directly inputting physical parameters, which is significant. Although, inputting physical parameters to guide prediction, perhaps by introducing a regularising term in the loss function, may be beneficial [56,57]. To demonstrate the learnt HFQ® dynamics, Figure 24 visualises an interpolation of stamping speeds from 50mm/s to 500mm/s for an arbitrary shrink corner with tight radii formed at two initial blank temperatures 350°C and 500°C. The thinning localisation reduces slowly with increasing speed, which is in good agreement with HFQ® literature [5,42]. Further, the smooth transition indicates an inexistence of overfitting, where network would otherwise memorise the training data. Moreover, all images were generated in real time, which facilitates a rapid identification of feasible processing conditions and blank shape for any given corner geometry design.

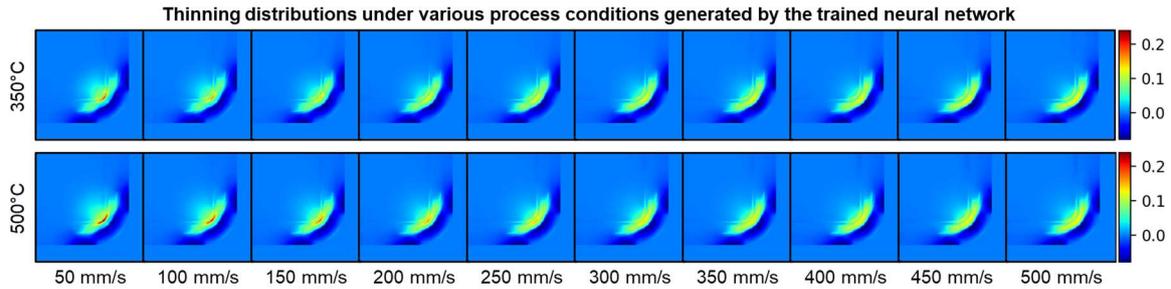

Figure 24 Interpolation of stamping speeds from 50mm/s to 500mm/s for initial blank temperatures of 350°C and 500°C. The smooth transition of thinning localisation to thinning uniformity shows the inexistence of overfitting and learnt HFQ® dynamics.

Although the training process of the networks is computationally expensive, approximately 2 days for the data generation and training steps combined, this process can be fully automated. Further, once trained networks are obtained, they can be employed to generate simulation results inexpensively and with high accuracy. The ability of the networks to provide rapid manufacturing feasibility assessment for any component geometry with any blank shape has profound consequences for the future of design for manufacture and early stage decision making for new, and existing, metal forming technologies.

# 7   Conclusions

HFQ® is a new non-isothermal hot stamping technology that can provide a means of forming complex shaped panel components from high strength aluminium alloys. However, the uptake of HFQ® is limited by its unfamiliarity among industrial designers. In response, this paper has presented a novel design support tool for HFQ® applications that can aid in early stage design and decision making. Based on the conducted research, the following conclusions can be summarised:

- A novel application of deep convolutional neural networks (CNNs) was presented for evaluating the manufacturing feasibility of components formed through the HFQ® process. The network architecture was tailored to extract geometric features from input images of design choices and generate corresponding forming responses under HFQ® conditions.

- In order to address the lack of data in the literature, novel algorithms were developed to automatically process CAD geometries and FE simulation results into image representations for network training and evaluation.

- Using the newly developed algorithms, a new dataset for shrink corners formed under non-isothermal HFQ® conditions was developed which contains a wide range of forming responses.

- Full thinning and displacement fields were predicted by the CNNs inexpensively and with high accuracy. Given a designed candidate geometry, it's manufacturing feasibility through HFQ® technology could be determined at early component design stages.

- In particular, the CNNs were able to capture non-trivial forming behaviours, such as material deformation of a 3D as-formed component, thinning localisation, and wrinkling.

The proposed methodology can be implemented in industrial settings to allow component designers to become familiar with the design limitations of the new HFQ® process without underlying knowledge of the process intricacies. In this way, component redesign for HFQ®, even for complex shapes, can be quickly identified and corrected at the onset of a design process. In order to allow for the discovery of new component designs which exploit the lightweighting potential of HFQ®, future research work will focus on the implementation of the proposed networks in design optimisation frameworks.

## Data Availability

The data that support this study is available from the corresponding author upon reasonable request.

## Acknowledgments

The authors thank the funding support by Impression Technologies Ltd, UK Engineering and Physical Sciences Research Council, Shougang Research Institute of Technology, and the Chinese Scholarship Council. Software and technical support from Rajab Said and Mustapha Ziane from ESI Group is also gratefully acknowledged. HFQ® is a registered trademark of Impression Technologies Ltd.

# Appendix

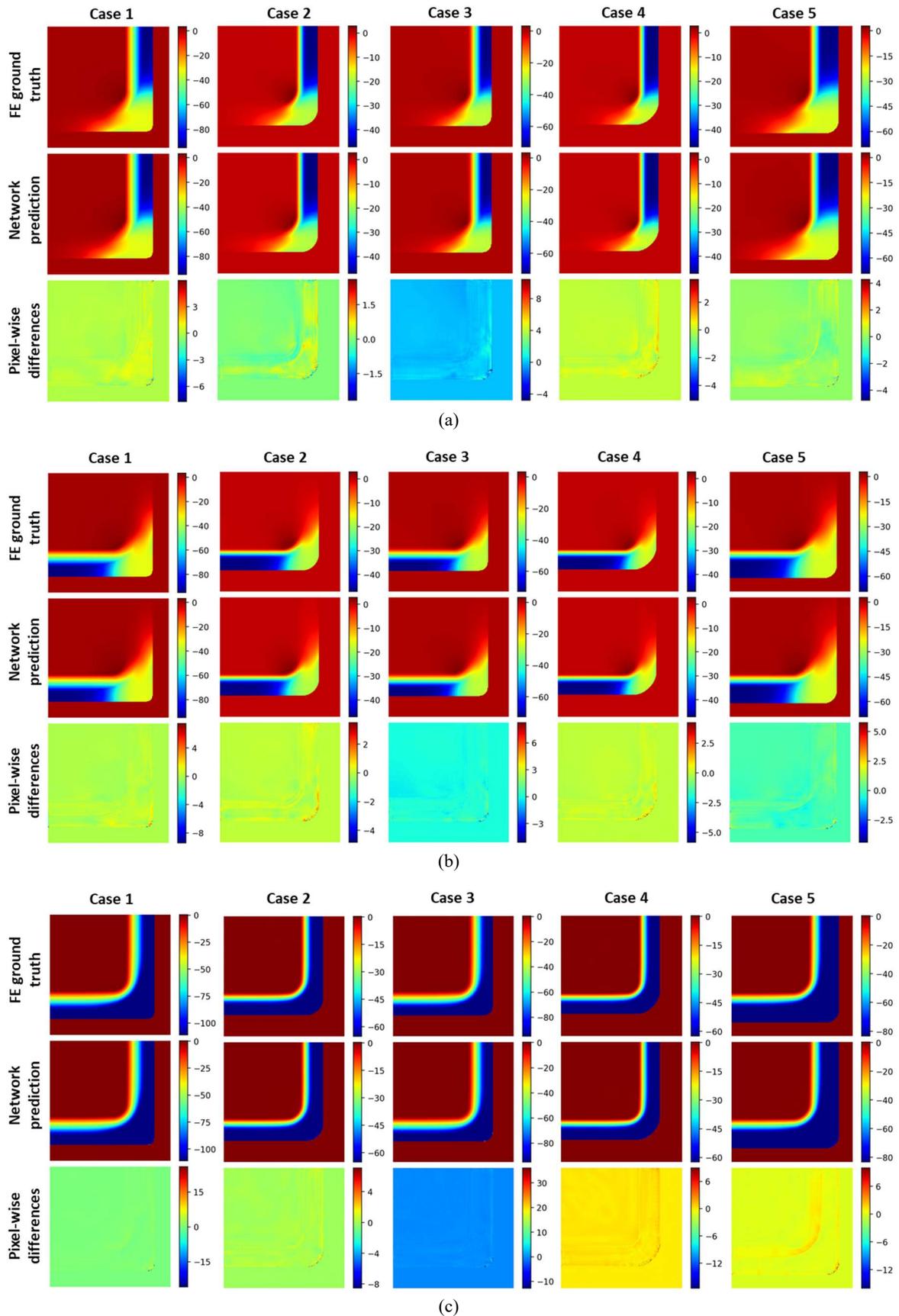

Figure A.1 Comparison of ground truth with network prediction for (a) X, (b) Y and (c) Z displacement distributions in mm and pixel wise differences for five random test set cases. Contours plotted on undeformed blank geometry.

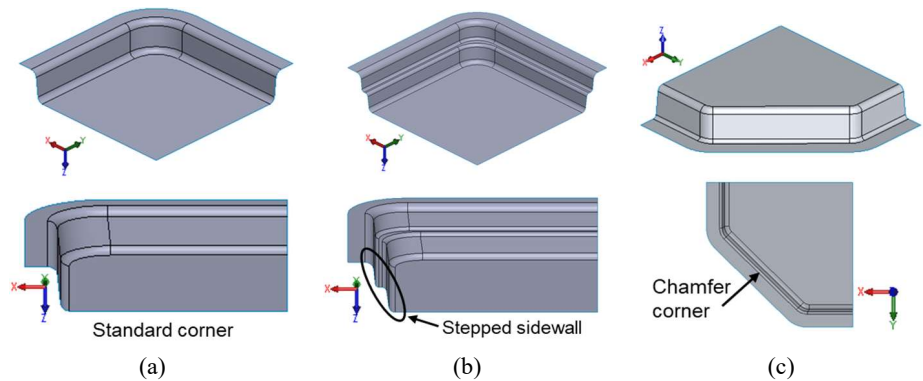

<div align="center">(a)          (b)          (c)</div>

Figure A.2 Shrink corner variants: (a) standard corners used to train the presented networks and (b) and (c) more complex design features typically represented by many CAD dimensions.

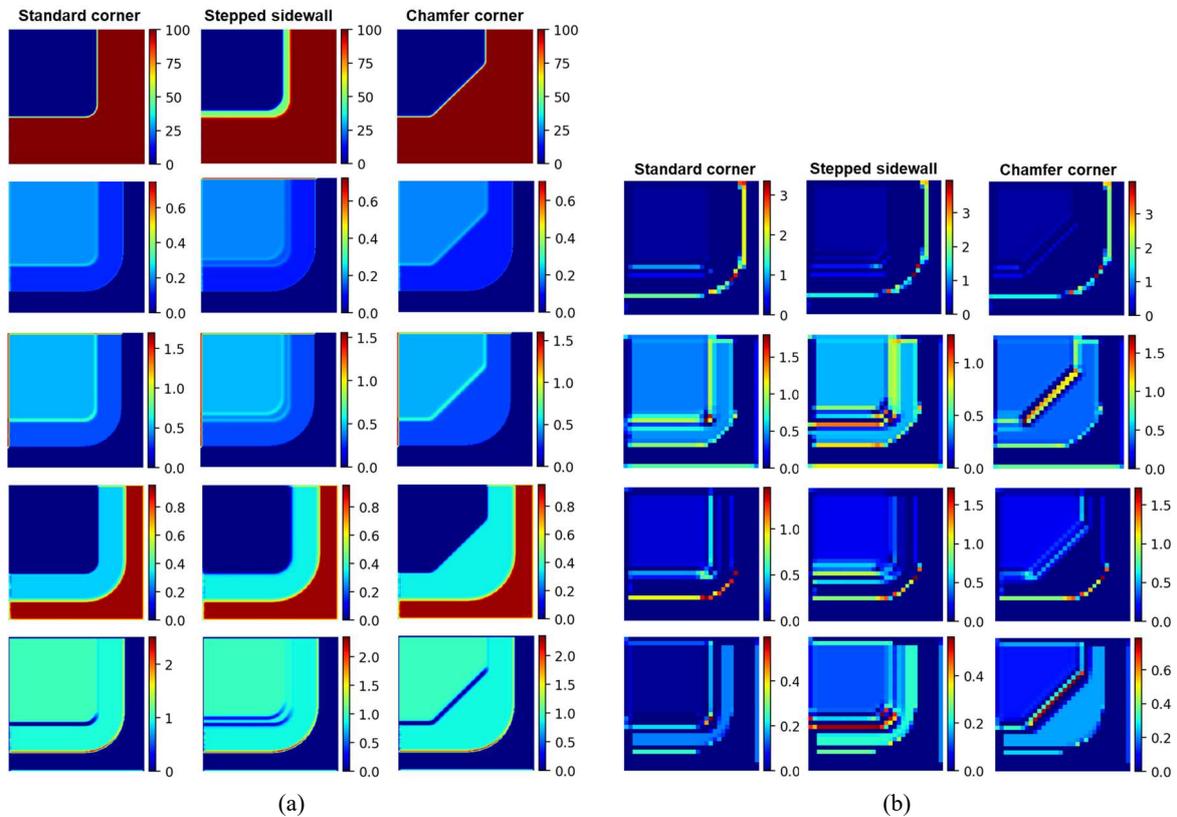

<div align="center">(a)                  (b)</div>

Figure A.3 Visualisation of encoder feature maps for different corner types, generated from a network trained only on images of standard corners: (a) Image representation of geometries from Figure (top row) with four random encoder feature maps from layer E1 (all other rows) for each corner design, (b) four random encoder feature maps from layer E4 for each corner. Same blank shape image used for each corner design.